# Quantum Nucleodynamics

Norman D. Cook
*Department of Informatics, Kansai University,*
*Takatsuki, Osaka, 569-1095 Japan*
*cook@res.kutc.kansai-u.ac.jp*

The core ideas underlying a quantitative, bottom-up theory of nuclear structure, i.e., quantum nucleodynamics, are introduced. The replacement of the fictitious "mean-field" approximation of the nuclear force with the empirically-known nuclear potential-well is the essential first step. From there, calculation of short-range nucleon-nucleon effects can be achieved on the basis of a unique lattice representation of nuclear quantum space, as introduced by Wigner in 1937. The lattice reproduces all of the ***n***-shells and ***j***- and ***m***-subshells of the well-established independent-particle model on a geometrical basis, and represents a return to a comprehensible, coordinate-space depiction of the atomic nucleus. Multiple regression results for the binding energies of 273 stable isotopes using 1$^{st}$-3$^{rd}$ neighbor interactions in the lattice are presented as "in principle" demonstration of the utility of the lattice. Noteworthy is the fact that all near-neighbor nuclear binding effects fall between -1.2 and 3.0 MeV.

PACS numbers: 20. Nuclear physics 21. Nuclear structure

## Introduction

"Quantum nucleodynamics" (QND) is a phrase that was used sporadically in the 1950s to describe the intended formalization of nuclear structure theory along the lines of quantum electrodynamics (QED). Unfortunately, despite the early development of a quantum mechanical foundation for nuclear theory (in the form of the independent-particle model, IPM), the nuclear version of QED turned out to be "so difficult that no one has ever been able to figure out what the consequences of the theory are" [1] and the promise of a unified, quantitative explanation of the atomic nucleus has not been realized.

The principal problem still lies in reconciling the diverse properties of nuclei with what is experimentally known about the "strong" nuclear force. Unlike QED, a fundamental theory of nucleon-nucleon interactions has not been established, despite the fact that realistic, ad hoc nuclear potentials have been devised on the basis of nucleon-nucleon scattering experiments. Already by the early 1960s, most theorists had turned their attention to high-energy *particle* physics in the hope that a quantitative theory of the strong force would clarify the puzzles of nuclear structure. Such efforts led successfully to the development of quantum chromo-dynamics (QCD), but have thus far failed to resolve the issues of either the nuclear force or nuclear structure itself. Meanwhile, the enticing QND phrase has been effectively abandoned and is rarely even mentioned in the physics literature (e.g., [1, 2]).

Here, I show that: (i) the foundations of a quantitative QND are already known and (ii) fully consistent with the well-established nuclear IPM. Furthermore, I argue that (iii) the development of QND awaits a focused research effort to complete the unfinished business of low-energy nuclear structure theory.

## Quantum Electrodynamics

At the onset of quantum theory, several profound philosophical debates concerning its interpretation were initiated and continue to be the source of controversy. Fortunately, practicing physicists can rely on the mathematical formalism of quantum mechanics to predict a broad range of atomic and nuclear phenomena. In that respect – and regardless of the philosophical issues of interpretation – there is no doubt that (i) quantum theory is fundamentally correct and, moreover, that (ii) its most precise applications are currently found in QED. Notably, despite differences of opinion concerning quantum philosophy (the collapse of the wave function, the implications of the uncertainty principle, the wave-particle duality, the stochastic/ deterministic nature of reality, etc.), there are today virtually no dissenting opinions concerning the unprecedented precision of QED – "the jewel of physics" [3]. As a quantitative theory that allows for an understanding of the absorption and emission of photons in terms of the transitions of electrons from one quantal state to another, QED remains unchallenged.

In spite of the technical complexity of quantum mechanics, in general, and QED, in particular, the conceptual simplicity of atomic theory can be illustrated as in Figure 1. As first understood by Niels Bohr in the 1920s, for a hydrogen-like atom in which there is one electron orbiting around a central nucleus containing Z charges, the entire set of excited states, their transitions and light spectra can be calculated on the basis of quantum theory (Figure 1A). Adding a second electron introduces electron-electron effects that can be computed, and further electrons introduce screening effects that must be handled on an *ad hoc* basis, but the fully developed theory of atomic structure remains qualitatively accurate and, with suitable parameter adjustments, quantitatively precise (Figure 1B).

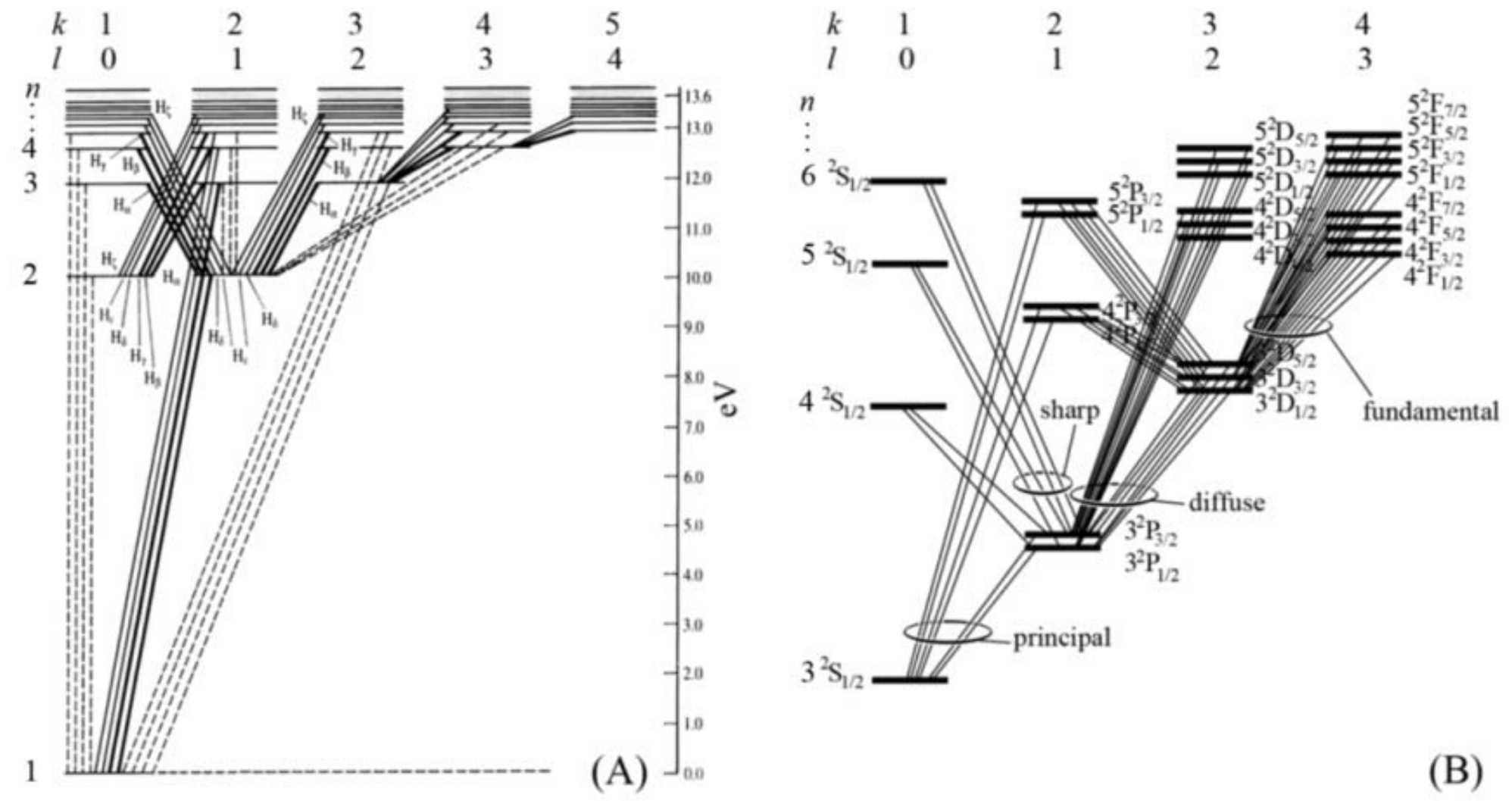

**Figure 1:** (A) The energy states of hydrogen – all of which can be calculated in quantum mechanics. (B) Energy levels and allowed one-electron transitions of the sodium atom. Essentially all excited-state atomic data can be understood on a similar basis.

| Quantum Numbers | K | L | | | | M | | | | | | | | |
|---|---|---|---|---|---|---|---|---|---|---|---|---|---|---|
| *n* | 1 | 2 | | | | 3 | | | | | | | | |
| *l* | 0 | 0 | 1 | | | 0 | 1 | | | 2 | | | | |
| | s | s | p | | | s | p | | | d | | | | |
| $m_l$ | 0 | 0 | -1 | 0 | 1 | 0 | -1 | 0 | 1 | -2 | -1 | 0 | 1 | 2 |
| $m_s$ | ⇧⇩ | ⇧⇩ | ⇧⇩ | ⇧⇩ | ⇧⇩ | ⇧⇩ | ⇧⇩ | ⇧⇩ | ⇧⇩ | ⇧⇩ | ⇧⇩ | ⇧⇩ | ⇧⇩ | ⇧⇩ |
| Closed (sub)Shells | K | (L1) | L2 | | | (M1) | M2 | | | (M3) | | | | |
| Number of States | 2 | 2 | 6 | | | 2 | 6 | | | 10 | | | | |
| Total Electrons | 2 | 10 | | | | 18 | | | | 36 | | | | |

| Quantum Numbers | N | | | | | | | | | O | | | |
|---|---|---|---|---|---|---|---|---|---|---|---|---|---|
| *n* | 4 | | | | | | | | | 5 | | | |
| *l* | 0 | 1 | | | 2 | | | | | 0 | 1 | | |
| | s | p | | | d | | | | | s | p | | |
| $m_l$ | 0 | -1 | 0 | 1 | -2 | -1 | 0 | 1 | 2 | 0 | -1 | 0 | 1 |
| $m_s$ | ⇧⇩ | ⇧⇩ | ⇧⇩ | ⇧⇩ | ⇧⇩ | ⇧⇩ | ⇧⇩ | ⇧⇩ | ⇧⇩ | ⇧⇩ | ⇧⇩ | ⇧⇩ | ⇧⇩ |
| Closed (sub)Shells | (N1) | N2 | | | (N3) | | | | | (O1) | O2 | | |
| Number of States | 2 | 6 | | | 10 | | | | | 2 | 6 | | |
| Total Electrons | | | | | 54 | | | | | | | | |

| Quantum Numbers | N | | | | | | | O | | | | | P | | | |
|---|---|---|---|---|---|---|---|---|---|---|---|---|---|---|---|---|
| *n* | 4 | | | | | | | 5 | | | | | 6 | | | |
| *l* | 3 | | | | | | | 2 | | | | | 0 | 1 | | |
| | f | | | | | | | d | | | | | s | p | | |
| $m_l$ | -3 | -2 | -1 | 0 | 1 | 2 | 3 | -2 | -1 | 0 | 1 | 2 | 0 | -1 | 0 | 1 |
| $m_s$ | ⇧⇩ | ⇧⇩ | ⇧⇩ | ⇧⇩ | ⇧⇩ | ⇧⇩ | ⇧⇩ | ⇧⇩ | ⇧⇩ | ⇧⇩ | ⇧⇩ | ⇧⇩ | ⇧⇩ | ⇧⇩ | ⇧⇩ | ⇧⇩ |
| Closed (sub)Shells | (N4) | | | | | | | (O3) | | | | | (P1) | P2 | | |
| Number of States | 14 | | | | | | | 10 | | | | | 2 | 6 | | |
| Total Electrons | 86 | | | | | | | | | | | | | | | |

**Table 1**: The full set of ***n***-, ***l***-, $m_l$- and $m_s$-quantal states of the first 86 electrons (after [4]). The "spherical harmonics" of electron orbitals makes quantum mechanics geometrically complex, but its underlying simplicity is based on the integer relationships among the quantum numbers.

Quantum mechanical results concerning the light spectra are the most spectacular, because theoretical predictions can be compared directly with experimental data. The central idea that makes prediction possible is that there is an underlying quantal "texture" of all possible electron states. As illustrated in Figure 1, any atomic state is a specific configuration of electrons with known ***n***, ***l*** and ***m*** quantum numbers that are used in the calculation of photon energies and of the allowed and forbidden transitions. The wave function that specifies the relationships among the quantum numbers is:

$$\Psi_{n,l,m} = R_{n,l}(r)\, Y_{m,l}(\theta, \phi) \qquad \text{Eq. 1}$$

Assuming that a central Coulomb potential binds electrons to the nucleus, Schrödinger was able to formulate a wave-equation for calculating electron energies, as in Eq. 2.

$$\Psi_{n,l,m} = H(q, p)\, R_{n,l}(r)\, Y_{m,l}(\theta, \phi) \qquad \text{Eq. 2}$$

The permutations of ***n***, ***l*** and ***m*** – and their dual occupancy with spin-up and spin-down electrons (***s***) provides the basic theoretical framework for determining the energy states of electrons (Eqs. 3-6, Table 1), and consequently light spectra. As stated in all textbooks on atomic theory, ***n***, ***l***, $m_l$ and $m_s$ can take certain integer or half-integer values:

$$n = 1, 2, 3, 4, \ldots \qquad \text{Eq. 3}$$
$$l = 0, 1, 2, \ldots, n\text{-}1 \qquad \text{Eq. 4}$$
$$m_l = -l, \ldots, -2, -1, 0, 1, 2, \ldots, l \qquad \text{Eq. 5}$$
$$m_s = s = 1/2, -1/2 \qquad \text{Eq. 6}$$

Based on the occupancy of electrons in the shells and subshells of Eqs. 1-6, it became possible to explain the length of the periods in the Periodic Table of the elements, and that theoretical achievement was a decisive factor in establishing quantum theory as the core explanation of atomic structure. Say what one will about the philosophical controversies in quantum theory, today the pattern of electron states (Table 1) is: (i) the bedrock of atomic theory, (ii) the foundation upon which an understanding of the Periodic Table is built, and (iii) the conceptual framework for essentially all of modern chemistry.

## Quantum Nucleodynamics

Although QED techniques were adapted to the problems of nuclear theory in the 1950s, they did *not* lead to a rigorous QND, primarily because the fundamental force holding nuclei together remained uncertain. As a consequence, nuclear structure theory did *not* progress to a quantitative explanation of the properties of nuclear ground- and excited-states, and has remained a realm of diverse, inherently imperfect and mutually-contradictory "models" [5]. Is that the inevitable fate of nuclear theory, or is there a prospect for a nuclear version of QED? More precisely, is it or is it not possible to employ a realistic nuclear potential-well (of the Paris, Bonn, Idaho or Argonne type), in order to make quantitative predictions about nuclear structure that are comparable to those already possible with regard to atomic structure?

Although nuclear structure and atomic structure clearly differ in terms of the forces holding these systems together, they are strikingly *similar* in terms of the wave-function description of the quantal states of their constituent fermions. That puzzle was tentatively resolved in the 1940s by assuming that the extremely short-range nuclear potential-well can be treated as a long-range "mean-field" – roughly analogous to the Coulomb potential felt by atomic electrons. That daring postulate led directly to the modern nuclear IPM, similar to the already-established atomic IPM, and provided indirect support for the idea of a nuclear interior with rather weakly-interacting nucleons. The presumed "independence" of both electrons and nucleons meant that local fermion-fermion interactions could be treated as minor perturbations. In the case of atomic theory, the assumption of a central potential-well for electrons was well justified in terms of the small Compton radius of the electron (r~2.8 fm) relative to the large atomic radius (r~100,000 fm or 1.0 Angstrom), but the same assumption in nuclear theory was inherently controversial because the nucleon radius (~0.9 fm) [6] is much larger relative to the nuclear radius (r<6 fm).

During the 1940s and 1950s while experimentalists successfully measured the ground- and excited-states of many thousands of isotopes, theorists labored to explain those states using various models, but without notable success. Specifically, it turned out that parameter sets for the nuclear force acting among all $A^2$ nucleon-pairs were unique to each particular isotope. Such isotope-by-isotope manipulation of nuclear force parameters has been described by shell model practitioners (e.g., [7]) as "rococo extravagance," but there has been no viable alternative for explaining nuclear structure under the still-questionable assumptions of effective mean-field theory.

Despite unresolved questions concerning the strong force, what came as a pleasant surprise already in the early 1950s was that, *similar* to atomic theory, every nucleon can be described by a unique set of quantum numbers, and the quantal state of the nucleus as a whole can be calculated from the summed states of its "independent" constituents. What is remarkably *different* is that the atomic system is a low-density "gas," whereas the nuclear system is an extremely high-density substance whose macroscopic properties had been, prior to the emergence of the shell model, well described in the liquid-drop model (LDM). The theoretical assumptions and parameters that have been invented to rationalize how analogous central potential-wells might explain both high- and low-density systems are discussed at length in the textbooks and the simultaneous existence of both gas and liquid properties is widely believed to be the inevitable reality underlying nuclear structure. On *a priori* grounds, it is difficult to know whether modern nuclear structure theory is better characterized as "rococo extravagance" or the "hard-headed science" of a highly complex object, but it is relevant to note that at least one alternative approach has emerged that does not rely on the simultaneous acceptance of both liquid-phase and gaseous-phase models. That is, a solid-phase model of the nucleus that is dominated by local, nearest-neighbor nucleon-nucleon interactions (similar to those assumed in the LDM) produces the same pattern of independent-particle quantal states (as postulated in the IPM), but without assuming a gaseous nuclear texture.

The foundations of the solid-phase model were laid in Eugene Wigner's Nobel Prize winning publications from the 1930s [8], and later developed by others [9-35]. The key insight, *explicitly stated by Wigner in 1937,* is that the quantal symmetries of nucleon eigenvalues correspond to the symmetries of a face-centered cubic (fcc) lattice: "the quantum numbers are all half-integers [whose] combinations form a face centered lattice…" ([8]). Wigner himself discussed nuclear states in terms of an abstract, multidimensional "momentum space," but all subsequent developments of the lattice model of nuclear structure have been in relation to coordinate space, i.e., 3D geometry.

The geometry of the nuclear lattice model has been described in detail many times (e.g., 24, 25, 27), but is often misunderstood as an attempt to return to pre-modern ideas about nuclear structure. In fact, whether a lattice, a liquid or a diffuse gas, the nucleus is a quantum mechanical object that defies common sense in many respects. With the inevitable, first impression of platonic solids, the lattice structures that accurately represent nuclear symmetries have been unfairly dismissed as wrong-headed attempts to return to pre-quantum ideas to explain the quantum mechanical nuclear reality. Such criticisms notwithstanding, the motivation underlying studies of the lattice model has been the unambiguous fact that there is a mathematical identity between nuclear quantal states and the symmetries inherent to, uniquely, an fcc lattice of nucleons [8]. Questions of interpretation of that identity are controversial and unresolved, but the identity itself remains unchallenged.

From the perspective of the conventional nuclear IPM that has dominated nuclear structure theory since 1949, the lattice representation of nuclear symmetries might be dismissed as a "lucky coincidence" between the known quantal symmetries of the nucleus and the inevitable arithmetic regularities of crystal symmetries, but the converse view is worth considering: Could the low-density nuclear IPM fortuitously mimic the symmetries of a lattice of nucleons, rather than vice versa? In terms of the known spatial dimensions of nucleons and nuclei, the high-density lattice is a far more realistic approximation of the nuclear texture than a Fermi gas and a lattice reproduces most of the macroscopic nuclear properties for which the LDM is justifiably famous. Moreover, the development of a theory of nuclear structure on the basis of the *known* short-range nuclear force is far more straight-forward than creating *de novo* a theoretical "mean-field" that is required to rationalize a gaseous model.

It is unlikely that such general arguments about the gross properties of the nuclear models can lead to firm conclusions, but the unanticipated symmetries of quantum numbers that are contained within the fcc lattice – as pointed out by Wigner, and elaborated in many subsequent papers [e.g., 9~35] – provide the necessary motivation for examining the merits of a high-density, nuclear model that is distinct from the highly-successful shell model, the equally-successful LDM and the surprisingly useful cluster models.

**A. Theoretical Framework**

The identity between the IPM and the lattice (outlined below) implies that *every* nuclear state in the IPM has an analog in 3D coordinate space. *Every* transition of nucleons from one quantal state to another – explicable in terms of integral changes in the quantum numbers of the nuclear wave-equation – necessarily corresponds to a specific vector in nuclear lattice space. As a consequence, without resorting to a theoretically-uncertain nuclear mean-field, the quantum mechanics of the gaseous-phase IPM can be reconstructed within the lattice based on realistic, near-neighbor, nucleon-nucleon interactions. From a computational perspective, the most significant aspect of the lattice is that its inherent geometry can be exploited in the development of a fine-grained, realistic, local-interaction version of the IPM, that is, what might be considered as the structural foundations of QND.

In comparison to atomic theory, there are two factors that increase the complexity of the nuclear version of the wave-function. The first is that the nucleus contains two types of nucleon, protons and neutrons, that are distinguished in terms of the so-called isospin quantum number, ***i***. The second is the notion of the coupling of orbital angular momentum (***l***) with intrinsic angular momentum (***s***) – giving each nucleon a total angular momentum quantum value ($\boldsymbol{j=l\pm s}$). As a consequence, the nuclear wave-function has additional subscripts (Eq. 7) and a slightly more complex pattern of shell/subshell occupancy (Table 2).

$$\Psi_{n,j\,(l\pm s),m,i} = R_{n,j\,(l\pm s),i}(r)\;Y_{m,j\,(l\pm s),i}(\theta,\phi) \qquad \text{Eq. 7}$$

Despite the additional quantum numbers, the nuclear version of the wave-equation holds the same promise as that realized in atomic theory, insofar as it implies a finite set of energy states into which nucleons can come and go with the release or absorption of photons. The universally-acknowledged strength of the nuclear IPM (ca. 1950) lay in the fact that each nucleon in the model has a unique set of quantum numbers (***n, j, m, l, s, i***), as specified in Eq. 7 and summarized in Table 2. Using that foundation for describing individual nucleons, the IPM makes it possible to explain *nuclear* states as the simple summation of the properties of its "independent" *nucleons* (e.g., Figure 2). Historically, the predictions of the IPM with regard to total angular momentum and parity states ($J^\pi$) were great successes in the early 1950s and led to optimistic predictions about the impending development of a rigorous, quantitative QND theory of nuclear structure.

| Quantum Numbers | | | | | | | | | | | | | | | | | | | | | | | | | | | |
|---|---|---|---|---|---|---|---|---|---|---|---|---|---|---|---|---|---|---|---|---|---|---|---|---|---|---|---|
| | *n* | 0 | 1 | | | 2 | | | | | | 3 | | | | | | | | | | 4 | | | | | |
| | *l* | 0 | 1 | | 0 | 2 | | | 1 | | 0 | 3 | | | | 2 | | | 1 | | 0 | 4 | | | | | … |
| | *j* | 1/2 | 3/2 | | 1/2 | 5/2 | | | 3/2 | | 1/2 | 7/2 | | | | 5/2 | | | 3/2 | | 1/2 | 9/2 | | | | | … |
| | \|*m*\| | 1/2 | 3/2 | 1/2 | 1/2 | 5/2 | 3/2 | 1/2 | 3/2 | 1/2 | 1/2 | 7/2 | 5/2 | 3/2 | 1/2 | 5/2 | 3/2 | 1/2 | 3/2 | 1/2 | 1/2 | 9/2 | 7/2 | 5/2 | 3/2 | 1/2 | 7/2 |
| | *s* | ⇧⇩ | ⇧⇩ | ⇧⇩ | ⇧⇩ | ⇧⇩ | ⇧⇩ | ⇧⇩ | ⇧⇩ | ⇧⇩ | ⇧⇩ | ⇧⇩ | ⇧⇩ | ⇧⇩ | ⇧⇩ | ⇧⇩ | ⇧⇩ | ⇧⇩ | ⇧⇩ | ⇧⇩ | ⇧⇩ | ⇧⇩ | ⇧⇩ | ⇧⇩ | ⇧⇩ | ⇧⇩ | ⇧⇩ |
| Number of States | | 2 | 4 | | 2 | 6 | | | 4 | | 2 | 8 | | | | 6 | | | 4 | | 2 | 10 | | | | | … |
| (Semi)magic Numbers | | **2** | (6) | | **8** | (14) | | | (18) | | **20** | **28** | | | | (34) | | | (38) | | (40) | **50** | | | | | … |
| Total Nucleons | *i* | 4 | 12 | | 16 | 28 | | | 36 | | 40 | 56 | | | | 68 | | | 76 | | 80 | 100 | | | | | … |

**Table 2:** The quantum states of the first 100 nucleons in the IPM. As in atomic physics the sequence of the theoretical shells and subshells can be adjusted to explain the existence of closed shells at the "magic" numbers.

Unfortunately, as noted above, the IPM was based on the dubious assumption of a long-range nuclear potential well – implying a gaseous nuclear interior with "point" nucleons orbiting unimpeded inside the nuclear interior. Although the central attractive force in *atomic* physics – where the nucleus itself attracts the orbiting electrons – was well-founded and the electron, as a particle, is *small* relative to the atomic volume, similar assumptions in nuclear theory have turned out to be grossly and demonstrably incorrect. The impossibility of intranuclear nucleon orbiting was not known to be the case in the 1930s, when the nuclear "Fermi gas" model was first considered, but the experimental work of Hofstadter in the early 1950s (Nobel Prize in 1961) showed that both the proton and the neutron have hard-core particle structure and RMS diameters of ~1.7 fm [36] (recent experimental data summarized by Sick [6]). Since a center-to-center nearest-neighbor internucleon distance of only 2.026 fm reproduces the known nuclear density (0.170 nucleons/fm$^3$), it is neither true that nucleons can be realistically thought of as "points" nor true that they are free to "orbit" in the nuclear interior without interacting with nucleons in their immediate vicinity. To deal with those inconvenient facts, a huge industry of theoretical developments ensued in the 1950s to explain *post hoc* the surprising successes of the IPM, but that effort has not led to clarity concerning either the nuclear force or the multitude of known excited states.

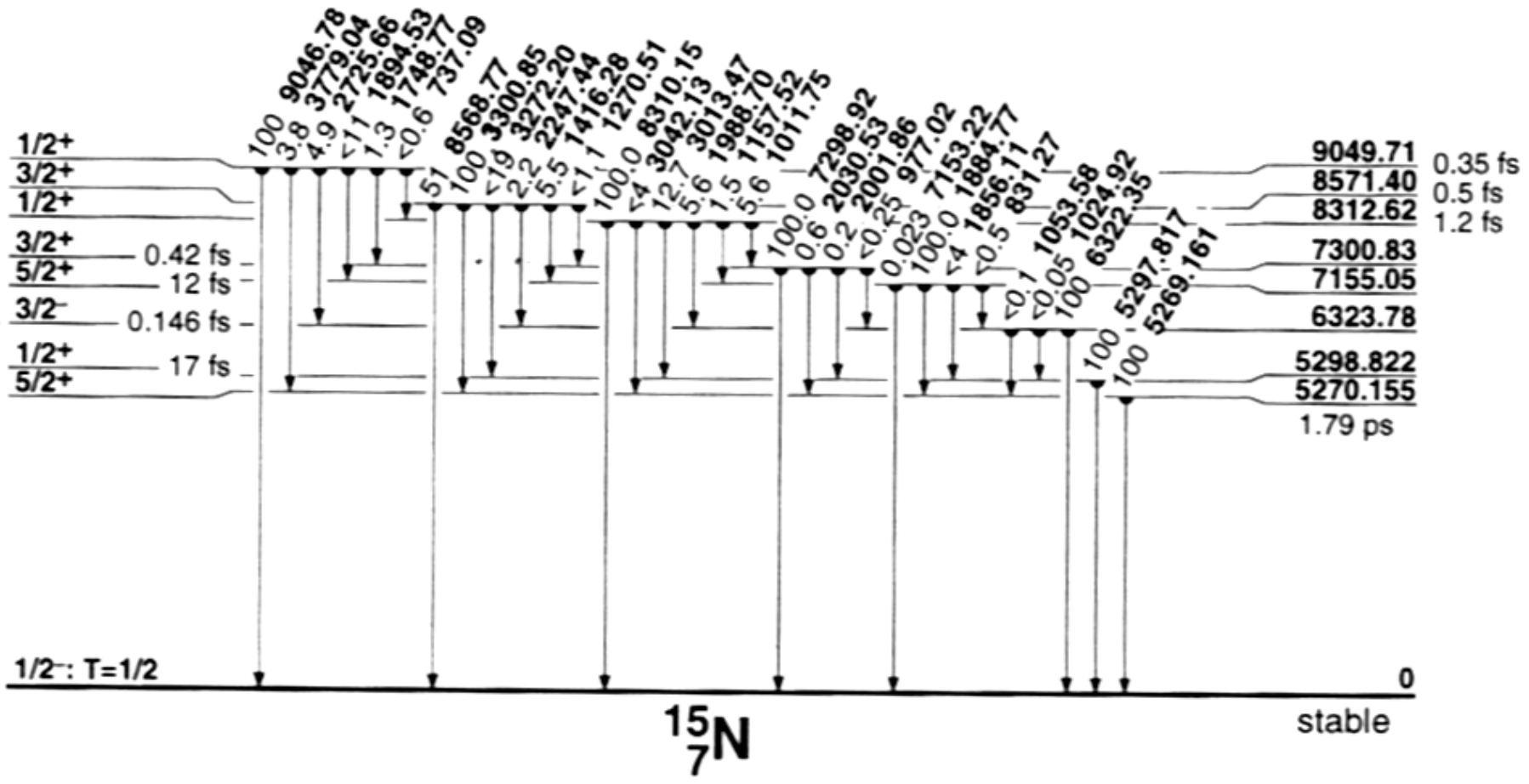

**Figure 2:** An example of the level of experimental detail known from nuclear spectroscopy [37]. The ***J***-values, parities, lifetimes, relative transition probabilities and energies of low-lying excited states have been measured for 2800+ isotopes. Although the ***J***-values and parities for $^{15}N$ (far left) are consistent with the IPM, the complex pattern of excitation energies, transition probabilities and lifetimes (center and far right) remains to be explained.

By the mid-1960s, nuclear structure theory had ossified into an utter paradox – an insoluble enigma where the nucleus is said to be simultaneously a dense-liquid and a diffuse-gas, and punctuated with alpha-particle clusters. Although it was a convoluted and unsatisfying theoretical story, it was also true that the pace of developments in nuclear weaponry, nuclear power and applications of nuclear isotopes made skepticism about the fundamentals of nuclear theory appear nonsensical. By the early 1950s, the nucleus had in fact been technologically "tamed" – but paradoxes unlike anything in atomic theory remained unresolved. Specifically, considerations of nuclear size, density and binding energies clearly demonstrated a high-density liquid-drop-like nuclear texture; considerations of alpha-decay, the binding energies of the small 4n-nuclei and high-energy multifragmentation results indicated the ubiquitous presence of alpha particles in large and small, stable and unstable nuclei; and considerations of nuclear spins, magnetic moments and parities suggested the reality of a nuclear gas with each "independent" nucleon having its own unique set of quantized properties. Theoretical self-consistency might be preferable, but the diverse models of traditional nuclear structure theory proved useful in explaining the empirical data. As a consequence of its quantum mechanical foundation, the IPM, rather than the liquid-drop or cluster models, became the centerpiece of nuclear structure theory and, ever since the late 1940s, theorists have struggled to justify the plausibility of a long-range "effective" nuclear mean-field of the nuclear force in a substance that appears to be a dense, chunky liquid.

By the early 1950s, the theoretical foundation of the nuclear IPM (Eq. 7) had been shown to be remarkably similar to that of atomic structure (Eq. 1), and the range of values that the quantum numbers in the nucleus can take (Eqs. 8-13) was found to be formally analogous to those of electron states (Eqs. 3-6):

$\boldsymbol{n}$ =0, 1, 2, 3, … Eq. 8
$\boldsymbol{l}$ = 0, 1, 2, …, (2$\boldsymbol{n}$)/2 Eq. 9
$\boldsymbol{j} = \boldsymbol{l} \pm \boldsymbol{s}$ = 1/2, 3/2, 5/2, …, (2$\boldsymbol{n}$+1)/2 Eq. 10
$\boldsymbol{m}$ = -$\boldsymbol{j}$, …, -5/2, -3/2, -1/2, 1/2, 3/2, 5/2, …, $\boldsymbol{j}$ Eq. 11
$\boldsymbol{s}$ =1/2, -1/2 Eq. 12
$\boldsymbol{i}$ =1, -1 Eq. 13

Together with the wave-equation itself, Eqs. 8-13 are essentially a concise statement of the established quantum mechanical behavior of the nucleus. In brief, both its IPM character and the "magic" numbers of the shell model can be obtained by manipulations of the nuclear shells and subshells (Table 2). The achievement of an over-arching theoretical framework within which the regularities of both atomic and nuclear structure could be explained was understood as demonstration that quantum mechanics lies at the heart of both systems. That argument – acknowledged by virtually every theorist involved in atomic or nuclear physics – remains a remarkable triumph of quantum theory.

Historically, Wigner's theoretical work on the quantal foundations of the IPM was immediately recognized as important – and eventually led to his sharing the Nobel Prize in physics with the inventors of the shell model. Largely ignored, however, was Wigner's lattice representation of nuclear states. In its modern form (including spin-orbit coupling), the lattice implies that nucleon quantum values can be used to deduce nucleon coordinates (x, y, z) (Eqs. 14-16):

$x = |2\boldsymbol{m}|(-1)^{(\boldsymbol{m}-1/2)}$ Eq. 14
$y = (2\boldsymbol{j}+1-|x|)(-1)^{(\boldsymbol{i}/2+\boldsymbol{j}+\boldsymbol{m}+1/2)}$ Eq. 15
$z = (2\boldsymbol{n}+3-|x|-|y|)(-1)^{(\boldsymbol{i}/2+\boldsymbol{n}-\boldsymbol{j}+1)}$ Eq. 16

And, conversely, the unique Cartesian coordinates for each nucleon can be used to define its quantal characteristics (Eqs. 17-23):

$\boldsymbol{n} = (|x| + |y| + |z| - 3) / 2$ Eq. 17
$\boldsymbol{l} = (|x| + |y| -2) / 2$ Eq. 18
$\boldsymbol{j} = (|x| + |y| -1) / 2$ Eq. 19
$\boldsymbol{m} = |x| *(-1)^{((x-1)/2)}/ 2$ Eq. 20
$\boldsymbol{s} = (-1)^{((x-1)/2)}/ 2$ Eq. 21
$\boldsymbol{i} = (-1)^{((z-1)/2)}$ Eq. 22
***parity*** = $\boldsymbol{\pi}$ = sign(x*y*z) Eq. 23

where x, y and z are the odd-integers that define an fcc lattice.

The significance of these extremely simple equations (Eqs.14~23) lies in the fact that, if we know the IPM (i.e., quantum mechanical) structure of a nucleus, then we also know its structure in coordinate space, and vice versa. The known pattern of quantum numbers and the occupancy of protons and neutrons in the ***n***-shells and in the ***j***- and ***m***-subshells are *identical* in both descriptions, but, in coordinate space, the abstract symmetries of the Schrödinger equation exhibit macroscopic geometrical symmetries, as well. The ***n***-shells and ***j***- and ***m***-subshells have spherical, cylindrical and conical symmetries, respectively, while ***s***- and ***i***-values produce orthogonal layering. Numerical examination of the symmetries in relation to the Cartesian coordinates shows the rigor and validity of Eqs. 14-23 (see also the Appendix) and the quantal texture of even the large nuclei can be easily visualized and verified using software designed for that purpose [31]. The mathematical isomorphism between quantum space and lattice space has been elaborated on in many publications over the past three decades, and recently summarized in a monograph [24]. The potential implications for the establishment of QND are, however, new and are outlined below.

**B. Qualitative QND**

The essential difference between the conventional IPM of nuclear structure theory and its lattice version lies in assumptions concerning the nuclear force. They both produce – identically – the same sequence and occupancy of quantal (***n, l, j, m, s, i***) states for any given number of protons and neutrons. In that respect, they are equivalent descriptions of the known independent-particle character of nuclei. There is, nonetheless, a huge difference between the two models. That is, in the conventional gaseous-phase IPM, there is no realistic possibility of calculating the local forces acting on nucleon ***a*** because nucleon ***a*** is assumed to be imbedded in the "mean-field" of all other nucleons orbiting within the nucleus and interacting with all other nucleons, ***b, c, d, …, z*** to varying and completely unknown degrees. In contrast, the same nucleon state in the lattice has an explicit set of local nucleon-nucleon interactions for 1st, 2nd and 3rd (etc.) nearest-neighbors, as implied by the lattice geometry. Computationally, that difference is significant because the lattice geometry is a (complex, but) tractable problem, whereas, in a gaseous phase, the computational problem is intractable for all medium and large nuclei. For a given isotope, either approach can be used *post hoc* to account for the experimentally-known set of excited states with specific energies, ***J***-values, parities and magnetic moments, but only the lattice version can state unambiguously that nucleon ***a*** with known quantal characteristics and known position within the lattice is a specific distance and orientation relative to nucleon ***b*** with its own quantal characteristics – and similarly for nucleons ***c*** through ***z***, and beyond.

In that regard, it can be said that, from the perspective of the lattice model, the conventional IPM is essentially correct in its independent-particle *description* of nuclear states. However, in assuming a gas-like random movement of point-particles and a mean-field nuclear force, the conventional model is inherently incapable of specifying the nature of the local nucleon-nucleon interactions. On the other hand, because of the lattice geometry, the lattice version of the IPM necessarily includes a complete specification of the nucleon-nucleon interactions that any particular nucleon imbedded in the lattice experiences. The nuclear lattice does not of course address issues of nucleon substructure or the interpretation of quantum theory itself, and the quantum "weirdness" of particle-waves remains an enigma in the lattice. Nevertheless, the nucleon lattice implies a geometrical substructure that is entirely absent in a nuclear "gas," but can be computationally exploited.

In effect, both the conventional and the lattice versions of the IPM can be used to *describe* any nuclear state and the transitions among the stable and excited states that are allowed, forbidden or "super-allowed." But drawing parallels between the experimental data and theory is not proof of either structure. Although both the gaseous-phase and solid-phase versions exhibit the same quantal descriptive powers, only the lattice makes possible the calculation of local 2-body nucleon-nucleon interactions. It is for this reason that the lattice version has the potential for becoming the basis for a rigorous QND theory of nuclear structure, whereas the gaseous version remains inherently "too difficult" (Feynman, 1963) – even with supercomputer assistance – and ends up with, at best, a vast array of adjustable parameters that must then be "fitted" to the empirical data.

**C. Quantitative QND**

At an abstract level, the traditional IPM and the lattice IPM are similar insofar as quantitative results that could be compared directly with experimental data require a full explication of the interactions of all nucleon pairs. In the conventional, gaseous IPM, the time-averaged distribution of nuclear matter must be calculated and the total point-to-point contribution of the assumed nuclear force then computed for all nucleon pairs. In principle, *de novo* calculation of nuclear binding energies, excited states, etc. is possible, but, even in the best-case scenario, the relative contributions of all nucleon interactions will require the *post hoc* weighting of the importance of the nucleon pairs. Precisely that kind of parameter adjustment is frequently employed in so-called *de novo* calculations, but ultimately means that computational results are not predictive, but rather are "retrodictive" accounting made through parameter adjustment guided by the experimental data.

In contrast, while the lattice version of the IPM implies a similar set of $A^2$ nucleon-nucleon permutations, the lattice geometry provides important constraints on permutations due to the fixed spatial dimensions of the lattice. The lattice IPM makes it theoretically possible to plug a model of the nuclear force directly into the lattice structures to obtain quantitative nuclear binding energies without *post hoc* adjustments. In principle, all possible permutations of ***n***, ***l***, ***j***, ***m***, ***s*** and ***i*** for each of the relevant distances between lattice sites can be tabulated and the strength of each solved using simultaneous equations techniques.

Currently, lattice QND is still semi-quantitative. Specifically, although a first-order description of nuclear ground-states in the lattice was achieved in the 1980s, a more detailed specification of nuclear force effects will require significant, modern-day computational resources. It is noteworthy that, as of today, calculation of all permutations of ***n***, ***l***, ***j***, ***m***, ***s*** and ***i*** at various internucleon distances within the lattice IPM can be calculated and parameter sets for nucleon-nucleon interactions can be explored. However, in order for the lattice model to become a rigorously quantitative QND theory (on a par with QED), it will be necessary to implement a realistic nuclear force within the lattice. In principle, that is already possible, but fine-tuning will require the efforts of many researchers.

**D. Multiple Regression**

The general case for thinking that the antiferro-magnetic fcc lattice is a coordinate-space representation of the traditional IPM has been presented above, but questions remain concerning how the lattice can be used to predict nuclear states. Results on nuclear binding energies have been obtained using multiple regression techniques and four such analyses are shown in Tables 3-6. In all four tables, a y-intercept of zero was chosen, in effect, forcing the regression line through the origin. That choice implicitly assumes that, in each analysis, there are no other relevant factors (that normally lead to a non-zero intercept), such that the chosen variables are adjusted within the multiple regression algorithm to account for the total binding energies. Depending on the number of variables, the assumption that other factors are irrelevant can be physically unrealistic, but allows for an evaluation of the overall ability of the assumed parameters in each analysis to explain nuclear binding energies.

The first analysis (Table 3) was done to determine the mean binding energy among nearest-neighbor nucleons in a close-packed lattice. For that purpose, 273 representative (stable and meta-stable) nuclei (A= 16~209) were built following the lattice definitions in Eqs. 13-23. Lattice structures were optimized such that (i) the lattice structure for each isotope had total spin and parity values ($J^\pi$) that were consistent with experimental values and, under that constraint, (ii) the structures had maximal nearest-neighbor nucleon-nucleon bonds and (iii) minimal Coulomb repulsion among protons. Note that the total Coulomb repulsion in the lattice was also calculated, and added to the experimental binding energy before solving for the mean binding energy per bond. Table 3 shows that a surprisingly small mean value of 2.733 MeV per nearest-neighbor nucleon-nucleon bond accounts for the total binding energy of nuclei. In other words, the BE per bond is relatively small and yet, since each nucleon has up to 12 nearest-neighbors, the total BE sums to several thousand MeV for the large nuclei.

Table 3: Multiple Regression on Nuclear Binding Energies with 1 Factor

| *Multiple Regression* | |
|---|---|
| R | 0.999947 |
| R2 | 0.999894 |
| Corr. R2 | 0.996217 |
| Std Dev | 15.10207 |
| No. Obs. | 273 |

| | *Coeff.* | *Std Dev* | *t* | *P-value* |
|---|---|---|---|---|
| Intercept | 0 | - | - | - |
| First Neighbors | 2.733 | 0.002 | 1599.216 | 0.000 |

Needless to say, the first analysis is not a realistic view of nuclear binding because a one-variable design ignores all possible higher-order effects on binding due to spin, isospin, etc. Nonetheless, it is of some interest that ~2.7 MeV is found to be the mean value per nearest-neighbor bond for nuclei ranging from $^{40}$Ca (2.779 MeV) to $^{104}$Pd (2.790 MeV) to $^{170}$Yt (2.693 MeV) to $^{209}$Bi (2.676 MeV). This theoretical result indicates that the lattice model has properties approximately similar to the LDM, and exhibits saturation of the nuclear binding force solely as a consequence of the binding energy among nearest-neighbors.

A slightly more realistic analysis was then carried out on the same set of nuclei, with consideration of the factor of distance among nucleons up to third nearest-neighbors. Table 4 shows that, on average, first-nearest neighbor interactions are attractive (2.697 MeV), second-nearest neighbors are repulsive (-1.172 MeV) and third-nearest neighbors are, on average, weakly attractive (0.592 MeV). Note that this analysis does not include the factor of isospin; all nucleons are treated as identical, with only the distance among lattice sites taken into consideration. It is noteworthy that the antiferromagnetic array of nucleons with alternating isospin layering implies attractive nearest-neighbors and repulsive second nearest-neighbors within layers, due to *magnetic* effects; third-nearest neighbors include both singlet and triplet pairs, and should therefore show weaker effects, as found in the regression analysis. In other words, the raw data obtained from the lattice build-up is consistent with the antiferromagnetic ordering of nucleons and indeed gives slightly improved statistical significance for the fit between experimental and theoretical BEs.

Typical of multiple regression analysis in general, R-values soon approach unity in all the analyses. The large R-values *cannot* be interpreted as indicating a high statistical likelihood that the values obtained for the nuclear force coefficients accurately reflect the physical character of the nuclear force, but R~0.9999 and relatively low standard deviations do indicate that there are few anomalies in the raw data (i.e., the data on the numbers of different types of nucleon-nucleon bonds within the lattice structures are self-consistent).

Table 4: Multiple Regression on Nuclear Binding Energies with 3 Factors

| *Multiple Regression* | |
|---|---|
| R | 0.99997 |
| R2 | 0.99995 |
| Corr. R2 | 0.99624 |
| Std. Dev. | 10.6808 |
| No. Obs. | 273 |

| | *Coeff.* | *Std. Dev.* | *t* | *P-value* |
|---|---|---|---|---|
| Intercept | 0 | - | - | - |
| 1st Neighbors | 2.697 | 0.034 | 78.226 | 0.000 |
| 2nd Neighbors | -1.172 | 0.072 | -16.236 | 0.000 |
| 3rd Neighbors | 0.592 | 0.047 | 12.505 | 0.000 |

The third analysis (Table 5) takes isospin and distance among the nucleons into consideration. Note that, in the antiferromagnetic fcc lattice with alternating isospin layers, first-neighbor and third-neighbor like-isospin combinations (PP1, NN1, PP3, NN3) are necessarily singlet pairs and second-neighbor like-isospin pairs (PP2 and NN2) are necessarily triplet pairs. There are no PN-pairs (PN2) at a distance of the second nearest-neighbor PP2 and NN2 pairs in the fcc lattice, while proton-neutron pairs (PN1 and PN3) include both singlet and triplet pairs. The relatively large coefficients for the first-neighbor bonds (and the correspondingly high-t and low-P values) are indication that nearest-neighbors have the strongest influence on nuclear binding, while second and third nearest-neighbors have weaker effects. All eight factors in this analysis were statistically significant ($p<0.009$) – again indicating that the raw data on bonding within the lattice are consistent with the known binding energies. Not surprisingly, with the addition of five new variables, the statistical significance of the fit is again found to improve.

Table 5: Multiple Regression on Nuclear Binding Energies with 8 Factors

| *Multiple Regression* | |
|---|---|
| R | 0.99999 |
| R2 | 0.99998 |
| Corr. R2 | 0.99620 |
| Std Dev | 7.48748 |
| No. Obs. | 273 |

| | *Coeff.* | *Std. Dev.* | *t* | *P-value* |
|---|---|---|---|---|
| Intercept | 0 | - | - | - |
| PP1 | 2.596 | 0.298 | 8.723 | 0.000 |
| PP2 | -0.867 | 0.226 | -3.828 | 0.000 |
| PP3 | 1.363 | 0.168 | 8.113 | 0.000 |
| NN1 | 1.762 | 0.217 | 8.133 | 0.000 |
| NN2 | -0.380 | 0.142 | -2.677 | 0.008 |
| NN3 | -0.605 | 0.117 | -5.191 | 0.000 |
| PN1 | 2.681 | 0.125 | 21.437 | 0.000 |
| PN3 | 0.796 | 0.072 | 11.046 | 0.000 |

Finally, a regression analysis that includes spin effects among the proton-neutron pairs is shown in Table 6. Note that first and third nearest-neighbor, like-isospin combinations are necessarily singlet-pairs ("s") with anti-parallel spins in the lattice (PP1s, PP3s, NN1s, NN3s) and second nearest-neighbor pairs are triplets ("t") with parallel spins. In contrast, first and third nearest-neighbor PN pairs include both singlet and triplet combinations (PN1s, PN1t, PN3s, PN3t). There are no second nearest-neighbor singlet or triplet PN2 pairs in the antiferromagnetic fcc lattice with alternating isospin layers.

In general, an increase in the number of regression coefficients leads to greater R-values and lower standard deviations. Although those statistics indicate increasingly better "fits" between the data and the model, the statistical strength of the individual coefficients gradually weakens, and confidence in the physical validity of the model can be strengthened only by increasing the size of the data set. Note that the standard deviation of ~7.44 MeV for the predictions of 273 nuclei with binding energies ranging from 127 to 1640 MeV is comparable to the conventional LDM. Moreover, it is noteworthy that all of the nuclear coefficients in the lattice model lend themselves to physically realistic interpretations in terms of a short-range nuclear force, and there is no need to append "shell corrections" or to include macroscopic adjustable parameters associated with the nuclear volume, deformation or surface area. In that regard, the lattice coefficients provide more physical insight than the variables used in the LDM because the latter variables, by definition, reflect collective properties of nuclei. In contrast, the coefficients in Table 6 reflect the character of specific types of nucleon-nucleon bonds and, in principle, can be tested for individual nuclei.

Table 6: Multiple Regression on Nuclear Binding Energies with 10 Factors

| *Multiple Regression* | |
|---|---|
| R | 0.99999 |
| R2 | 0.99998 |
| Corr. R2 | 0.99620 |
| Std Dev | 7.44021 |
| No. Obs. | 273 |

| | *Coeff.* | *Std.Dev.* | *t* | *P-value* |
|---|---|---|---|---|
| Intercept | 0 | - | - | - |
| PP1s | 2.715 | 0.301 | 9.035 | 0.000 |
| PP2t | -0.918 | 0.226 | -4.054 | 0.000 |
| PP3s | 1.436 | 0.170 | 8.449 | 0.000 |
| NN1s | 1.804 | 0.217 | 8.306 | 0.000 |
| NN2t | -0.332 | 0.145 | -2.284 | 0.020 |
| NN3s | -0.615 | 0.116 | -5.298 | 0.000 |
| PN1t | 3.025 | 0.447 | 6.772 | 0.000 |
| PN1s | 2.258 | 0.470 | 4.807 | 0.000 |
| PN3t | 1.208 | 0.195 | 6.209 | 0.000 |
| PN3s | 0.338 | 0.210 | 1.607 | 0.109 |

A similar analysis using the full set of permutations of all possible nucleon states is the next obvious step to take. Unfortunately, the large number of nucleon-nucleon combinations – each with distinct spin, isospin, ***j***- and ***m***-quantum values (Table 7) – makes statistical analysis impossible until the lattice data from many thousands of ground and excited states are included. Such analysis is, in principle, possible today, but will require a large-scale research effort, as was indeed undertaken for the development of the shell model in the 1950s.

Note that the 80 permutations listed in Table 7 are a much smaller set of permutations than are calculated for the Clebsch-Gordon coefficients. Only 80 coefficients are required in the lattice model because of the geometrical constraints of the lattice (e.g., there are no spin(1/2)-spin(7/2) combinations within the distance of 3rd-nearest-neighbors), whereas all possible spin permutations are possible in the IPM and must therefore be included in gaseous-phase model calculations.

The difference between the traditional IPM and the lattice IPM can be stated as a difference in the nature of the truncation needed for computational work. In the gaseous-phase IPM, there is no truncation that is fully justified, insofar as two nucleons with very different quantal characteristics may find themselves near to one another in coordinate space. In contrast, the deep relationship between quantum space and coordinate space in the lattice model implies a natural truncation for nucleon pairs that are distant within the lattice space. By excluding all nuclear force effects between nucleons that are 4th (or larger) nearest neighbors (i.e., separated by 4.0 fm or more), only the small contributions of very distant neighbors are neglected in the lattice model.

Whatever the final outcome of such statistical analysis, the essential merit of the lattice version of the IPM is that *all* two-body nucleon-nucleon interactions in a given nucleus can be specified in terms of nucleon quantum numbers (known from the conventional IPM) and *distances within the lattice*. By completely discarding the assumption of "intranuclear orbiting" of nucleons, each nucleon has a relatively small number of neighbors that can be expected to influence nuclear binding. Specifically, each nucleon in the close-packed lattice has a maximum of 12 nearest-neighbors, 6 second nearest-neighbors and 24 third nearest-neighbors. The effects among even further neighbors can, in principle, be calculated, but would likely be a small correction on total binding energies. In either case, the fixed dimensions of a lattice of nucleons with a core density of 0.17 nucleons/$fm^3$ mean that any realistic model of the nuclear force (Paris, Bonn, Argonne, Idaho, etc.) that makes use of those same parameters (internucleon distance, spin, isospin, ***j*** and ***m***) should be directly applicable within the lattice, and allow for quantitative predictions concerning both ground-state binding energies and excited states.

Table 7: The full set of nuclear force coefficients for the 80 permutations (***j, m, s, i***) among 1st (PP1, NN1, PN1), 2nd (PP2, NN2) and 3rd (PP3, NN3, PN3) nearest-neighbor nucleon-nucleon combinations, up to j=5/2 and m=5/2. All nearest-neighbor PP and NN combinations are singlet pairs (s), and all 2nd nearest-neighbor PP and NN combinations are triplet pairs (t), whereas PN1 and PN3 pairs include both singlet and triplet combinations. These coefficients correspond to a subset of the Clebsch-Gordon coefficients used in the conventional IPM.

| Lattice Definitions | Nuclear Force Coeff. | Lattice Definitions | Nuclear Force Coeff. |
|---|---|---|---|
| PP1$_{(j1/2-1/2)(m1/2-1/2)}$s | - | NN1$_{(j1/2-1/2)(m1/2-1/2)}$s | - |
| PP1$_{(j1/2-3/2)(m1/2-1/2)}$s | - | NN1$_{(j1/2-1/2)(m1/2-1/2)}$s | - |
| PP1$_{(j1/2-3/2)(m1/2-3/2)}$s | - | NN1$_{(j1/2-1/2)(m1/2-1/2)}$s | - |
| PP1$_{(j1/2-5/2)(m1/2-1/2)}$s | - | NN1$_{(j1/2-5/2)(m1/2-1/2)}$s | - |
| PP1$_{(j1/2-5/2)(m1/2-3/2)}$s | - | NN1$_{(j1/2-5/2)(m1/2-3/2)}$s | - |
| PP1$_{(j3/2-3/2)(m1/2-1/2)}$s | - | NN1$_{(j3/2-3/2)(m1/2-1/2)}$s | - |
| PP1$_{(j3/2-3/2)(m1/2-3/2)}$s | - | NN1$_{(j3/2-3/2)(m1/2-3/2)}$s | - |
| PP1$_{(j3/2-3/2)(m3/2-5/2)}$s | - | NN1$_{(j3/2-3/2)(m3/2-5/2)}$s | - |
| … | | | |
| PP2$_{(j1/2-1/2)(m1/2-1/2)}$t | - | NN2$_{(j1/2-1/2)(m1/2-1/2)}$t | - |
| PP2$_{(j1/2-3/2)(m1/2-1/2)}$t | - | NN2$_{(j1/2-1/2)(m1/2-1/2)}$t | - |
| PP2$_{(j1/2-3/2)(m1/2-3/2)}$t | - | NN2$_{(j1/2-1/2)(m1/2-1/2)}$t | - |
| PP2$_{(j1/2-5/2)(m1/2-1/2)}$t | - | NN2$_{(j1/2-5/2)(m1/2-1/2)}$t | - |
| PP2$_{(j1/2-5/2)(m1/2-3/2)}$t | - | NN2$_{(j1/2-5/2)(m1/2-3/2)}$t | - |
| PP2$_{(j3/2-3/2)(m1/2-1/2)}$t | - | NN2$_{(j3/2-3/2)(m1/2-1/2)}$t | - |
| PP2$_{(j3/2-3/2)(m1/2-3/2)}$t | - | NN2$_{(j3/2-3/2)(m1/2-3/2)}$t | - |
| PP2$_{(j3/2-3/2)(m3/2-5/2)}$t | - | NN2$_{(j3/2-3/2)(m3/2-5/2)}$t | - |
| … | | | |
| PP3$_{(j1/2-1/2)(m1/2-1/2)}$s | - | NN3$_{(j1/2-1/2)(m1/2-1/2)}$s | - |
| PP3$_{(j1/2-3/2)(m1/2-1/2)}$s | - | NN3$_{(j1/2-1/2)(m1/2-1/2)}$s | - |
| PP3$_{(j1/2-3/2)(m1/2-3/2)}$s | - | NN3$_{(j1/2-1/2)(m1/2-1/2)}$s | - |
| PP3$_{(j1/2-5/2)(m1/2-1/2)}$s | - | NN3$_{(j1/2-5/2)(m1/2-1/2)}$s | - |
| PP3$_{(j1/2-5/2)(m1/2-3/2)}$s | - | NN3$_{(j1/2-5/2)(m1/2-3/2)}$s | - |
| PP3$_{(j3/2-3/2)(m1/2-1/2)}$s | - | NN3$_{(j3/2-3/2)(m1/2-1/2)}$s | - |
| PP3$_{(j3/2-3/2)(m1/2-3/2)}$s | - | NN3$_{(j3/2-3/2)(m1/2-3/2)}$s | - |
| PP3$_{(j3/2-3/2)(m3/2-5/2)}$s | - | NN3$_{(j3/2-3/2)(m3/2-5/2)}$s | - |
| … | | | |
| PN1$_{(j1/2-1/2)(m1/2-1/2)}$s | - | PN3$_{(j1/2-1/2)(m1/2-1/2)}$s | - |
| PN1$_{(j1/2-3/2)(m1/2-1/2)}$s | - | PN3$_{(j1/2-3/2)(m1/2-1/2)}$s | - |
| PN1$_{(j1/2-3/2)(m1/2-3/2)}$s | - | PN3$_{(j1/2-3/2)(m1/2-3/2)}$s | - |
| PN1$_{(j1/2-5/2)(m1/2-1/2)}$s | - | PN3$_{(j1/2-5/2)(m1/2-1/2)}$s | - |
| PN1$_{(j1/2-5/2)(m1/2-3/2)}$s | - | PN3$_{(j1/2-5/2)(m1/2-3/2)}$s | - |
| PN1$_{(j3/2-3/2)(m1/2-1/2)}$s | - | PN3$_{(j3/2-3/2)(m1/2-1/2)}$s | - |
| PN1$_{(j3/2-3/2)(m1/2-3/2)}$s | - | PN3$_{(j3/2-3/2)(m1/2-3/2)}$s | - |
| PN1$_{(j3/2-3/2)(m3/2-5/2)}$s | - | PN3$_{(j3/2-3/2)(m3/2-5/2)}$s | - |
| … | | | |
| PN1$_{(j1/2-1/2)(m1/2-1/2)}$t | - | PN3$_{(j1/2-1/2)(m1/2-1/2)}$t | - |
| PN1$_{(j1/2-3/2)(m1/2-1/2)}$t | - | PN3$_{(j1/2-3/2)(m1/2-1/2)}$t | - |
| PN1$_{(j1/2-3/2)(m1/2-3/2)}$t | - | PN3$_{(j1/2-3/2)(m1/2-3/2)}$t | - |
| PN1$_{(j1/2-5/2)(m1/2-1/2)}$t | - | PN3$_{(j1/2-5/2)(m1/2-1/2)}$t | - |
| PN1$_{(j1/2-5/2)(m1/2-3/2)}$t | - | PN3$_{(j1/2-5/2)(m1/2-3/2)}$t | - |
| PN1$_{(j3/2-3/2)(m1/2-1/2)}$t | - | PN3$_{(j3/2-3/2)(m1/2-1/2)}$t | - |
| PN1$_{(j3/2-3/2)(m1/2-3/2)}$t | - | PN3$_{(j3/2-3/2)(m1/2-3/2)}$t | - |
| PN1$_{(j3/2-3/2)(m3/2-5/2)}$t | - | PN3$_{(j3/2-3/2)(m3/2-5/2)}$t | - |
| … | | | |

Stated in terms of research strategy, progress in explaining nuclear binding energies using the lattice model and realistic models of the nuclear force will require concerted computational efforts, but does not imply the need for quantum computing or memory-storage that remains many orders-of-magnitude beyond current capabilities. In other words, the computational problem of nuclear binding energies is entirely tractable within the constraints of a lattice of nucleons. While "old school" nuclear structure theorists may present *a priori* arguments concerning why a nuclear lattice model "should not work," the traditional approach in nuclear structure theory has been, to the contrary, to develop computational techniques to find out if model calculations are useful. In fact, unresolved questions concerning theoretical foundations notwithstanding, a great many nuclear structure models [5] have been found to be useful. As a consequence, as of the early 21$^{st}$ century, few nuclear structure theorists would be willing to state categorically that a liquid-phase or a gaseous-phase or a cluster model of nuclear structure "should not work," but, as a matter of fact, few theorists have made efforts in exploring the implications of a solid-phase lattice model.

## Conclusion

Subsequent to the (re)discovery of the fact that the internal symmetries of an fcc lattice reproduce the well-established symmetries of the conventional IPM, the lattice model has been developed to allow for predictions concerning primarily nuclear spins, parities and various macroscopic properties [9-35]. Arguments concerning the possible "unification" of the nuclear models and visualization of nuclear structure remain of peripheral interest, but a far more valuable step would be the development of techniques for making predictions of nuclear binding energies on the basis of the nuclear lattice and the empirical properties of the nuclear force, as known from nucleon-nucleon scattering experiments. In that respect, the establishment of a computational, lattice-based QND should be welcomed by theorists of all backgrounds and would essentially eliminate the need to choose a nuclear model before engaging in quantitative work: One would then "choose" quantum mechanics and then calculate the full set of two-body interactions implied by the lattice representation of the quantum mechanical IPM. Three-body and higher-order interactions among nucleons within the lattice might someday provide even greater precision, but the known lattice dimensions and symmetries already indicate a self-consistent, first-order description of nuclear states that is deducible solely from two-body interactions.

Given the mathematical identity between the gaseous-phase IPM and the lattice IPM, the theoretical situation in nuclear structure physics in the early 21$^{st}$ century is curiously similar to that in chemistry in the middle of the 19$^{th}$ century. In both fields a foundation of empirical findings was first established from painstaking laboratory work, where the primary data were masses and dissociation energies. Initially, 3D configurations of particles were *not* thought to be either realistic as depictions of the physical reality or useful as heuristics for theoretical study. The most notorious example of the neglect of geometrical considerations in chemistry concerns the benzene molecule. On the basis of experimental work, benzene had been determined to consist of 6 carbon atoms and 6 hydrogen atoms, $C_6H_6$. Kekulé proposed a hexagonal ring of carbons, but for *decades* the academic authorities in chemistry rejected all notions of molecular structure – both Kekulé's ring structure for benzene and van't Hoff's notion of tetra-hedral geometry for the carbon atom. Journal editors, such as A.W.H. Kolbe, famously argued that stereochemistry was "loose speculation parading as theory" indulged in by those with "no liking for exact chemical investigation" [38]. Eventually, of course, Kekulé became known as the father of modern stereochemistry, and three of his students, including van't Hoff, won Nobel Prizes in chemistry in the early 20$^{th}$ century, specifically for their insights concerning molecular geometry.

It is relevant to note that the rejection of notions of 3D structure in 19$^{th}$ century chemistry was totally unrelated to the subsequent philosophical quandaries of the interpretation of the uncertainty principle, the wave/particle duality or the collapse of the wave equation, etc. Indeed, quantum mechanics did not emerge until several decades later, but there was nonetheless, already in the mid-19$^{th}$ century, a strong reluctance among practicing chemists to "speculate" about spatial structure [39]. Understandably, perhaps, most chemists had their hands full with the issues of laboratory experimentation and were wary of the daunting complexity of the structural permutations implied by stereochemical considerations. Gradually and inevitably, the field of chemistry evolved into experimental and theoretical branches, and the theorists eventually came to understand the necessity of including the constraints of molecular geometry. Ultimately, the blanket dismissal of the complexities of 3D structure by "old school" laboratory chemists proved to be unfounded, and stereochemistry has of course become a mainstream topic in all aspects of chemistry, biochemistry and molecular biology.

In the early 21$^{st}$ century, nuclear physics has arrived at a similar fork in the road, where "old school" experimentalists maintain that there is no discernible spatial substructure inherent to the pattern of nuclear data, such as shown in Figure 2. In effect, they argue that there is no possible deconvolution of multiple, over-lapping wave-functions into structural subcomponents and that there is no geometry "hidden" behind the spectroscopic data. As a consequence, rather than "speculate" about 3D structure, many theorists are hopeful that current difficulties in understanding the complexities of many-body nuclear systems might eventually be overcome by supercomputing hardware/ software capabilities without addressing issues of nuclear geometry. Following in the footsteps of Niels Bohr, some have even argued that the very idea of nuclear substructure "violates" the collective nature of nuclei and, today, only a small minority is actively "speculating" on the internal substructure of nuclei. I would suggest that the lattice representation of the IPM indicates a possible way forward toward a quantitative QND that builds upon the robust successes of the traditional IPM, and supplements it with geometrical considerations.

In producing comprehensive lists of empirical data on all of the 2800+ known isotopes [37], it is arguably the case that experimentalists have completed their primary task, but significant theoretical efforts – comparable to those of the early stereochemists – are still required to clarify the meaning of the several million data points on the structure of the ground- and low-lying excited-states of nuclei. The geometrical permutations of a nucleus containing A-nucleons are of course numerous, but the constraints of the antiferromagnetic fcc lattice greatly reduce the structural possibilities relative to those in molecular stereochemistry.

Today, more than six decades after the triumph of the IPM, the field of nuclear structure physics is arguably on the brink of establishing the structural rules of QND. An understanding of the allowed and forbidden nuclear transitions and their energies on the basis of underlying structural principles would permit nuclear theory to move from the stage of *ad hoc* models with inherently un-interpretable parameter sets to a coherent quantum mechanical theory. But to take that step, it will be necessary to acknowledge, above all else, that the long-range, "effective" nuclear potential-well that has been the plaything of nuclear theorists since the late 1940s was a mistaken assumption that is not justified from what is known experimentally about the actual nuclear force and that has not, in fact, led to a comprehensive, self-consistent, quantitative theory of nuclear structure.

---

# Appendix

| A | B | C | D | E | F | G | H | I | J | K | L | M | N | O | P |
|---|---|---|---|---|---|---|---|---|---|---|---|---|---|---|---|
| | | (Keyboard Input) | | | Quantum Numbers and Parities Calculated | | | | | | | Lattice Sites Calculated from the | | | X,Y,Z |
| Nucleon Sequence | | FCC Lattice Sites | | | from the Lattice Sites (Eqs. 16-22) | | | | | | | Quantum Numbers (Eqs. 13-15) | | | coords. |
| Proton No. | Neutron No. | X | Y | Z | *n* | *L* | *j* | *m* | *s* | *i* | *parity* | X | Y | Z | Check |
| | 1 | 1 | -1 | -1 | 0 | 0 | 1/2 | 1/2 | 1/2 | -1 | 1 | 1 | -1 | -1 | ok |
| 1 | | 1 | 1 | 1 | 0 | 0 | 1/2 | 1/2 | 1/2 | 1 | 1 | 1 | 1 | 1 | ok |
| | 2 | -1 | 1 | -1 | 0 | 0 | 1/2 | -1/2 | -1/2 | -1 | 1 | -1 | 1 | -1 | ok |
| 2 | | -1 | -1 | 1 | 0 | 0 | 1/2 | -1/2 | -1/2 | 1 | 1 | -1 | -1 | 1 | ok |
| | 3 | 1 | 3 | -1 | 1 | 1 | 3/2 | 1/2 | 1/2 | -1 | -1 | 1 | 3 | -1 | ok |
| 3 | | 3 | -1 | 1 | 1 | 1 | 3/2 | -3/2 | -1/2 | 1 | -1 | 3 | -1 | 1 | ok |
| | 4 | -1 | -3 | -1 | 1 | 1 | 3/2 | -1/2 | -1/2 | -1 | -1 | -1 | -3 | -1 | ok |
| 4 | | -1 | 3 | 1 | 1 | 1 | 3/2 | -1/2 | -1/2 | 1 | -1 | -1 | 3 | 1 | ok |
| | 5 | 3 | 1 | -1 | 1 | 1 | 3/2 | -3/2 | -1/2 | -1 | -1 | 3 | 1 | -1 | ok |
| 5 | | -3 | 1 | 1 | 1 | 1 | 3/2 | 3/2 | 1/2 | 1 | -1 | -3 | 1 | 1 | ok |
| | 6 | -3 | -1 | -1 | 1 | 1 | 3/2 | 3/2 | 1/2 | -1 | -1 | -3 | -1 | -1 | ok |
| 6 | | 1 | -3 | 1 | 1 | 1 | 3/2 | 1/2 | 1/2 | 1 | -1 | 1 | -3 | 1 | ok |
| | 7 | -1 | 1 | 3 | 1 | 0 | 1/2 | -1/2 | -1/2 | -1 | -1 | -1 | 1 | 3 | ok |
| 7 | | 1 | 1 | -3 | 1 | 0 | 1/2 | 1/2 | 1/2 | 1 | -1 | 1 | 1 | -3 | ok |
| | 8 | 1 | -1 | 3 | 1 | 0 | 1/2 | 1/2 | 1/2 | -1 | -1 | 1 | -1 | 3 | ok |
| 8 | | -1 | -1 | -3 | 1 | 0 | 1/2 | -1/2 | -1/2 | 1 | -1 | -1 | -1 | -3 | ok |
| | 9 | 3 | -3 | -1 | 2 | 2 | 5/2 | -3/2 | -1/2 | -1 | 1 | 3 | -3 | -1 | ok |
| 9 | | 3 | 3 | 1 | 2 | 2 | 5/2 | -3/2 | -1/2 | 1 | 1 | 3 | 3 | 1 | ok |
| | 10 | -3 | 3 | -1 | 2 | 2 | 5/2 | 3/2 | 1/2 | -1 | 1 | -3 | 3 | -1 | ok |
| 10 | | -3 | -3 | 1 | 2 | 2 | 5/2 | 3/2 | 1/2 | 1 | 1 | -3 | -3 | 1 | ok |
| | 11 | 5 | -1 | -1 | 2 | 2 | 5/2 | 5/2 | 1/2 | -1 | 1 | 5 | -1 | -1 | ok |
| 11 | | 5 | 1 | 1 | 2 | 2 | 5/2 | 5/2 | 1/2 | 1 | 1 | 5 | 1 | 1 | ok |
| | 12 | -5 | 1 | -1 | 2 | 2 | 5/2 | -5/2 | -1/2 | -1 | 1 | -5 | 1 | -1 | ok |
| 12 | | 1 | 5 | 1 | 2 | 2 | 5/2 | 1/2 | 1/2 | 1 | 1 | 1 | 5 | 1 | ok |
| | 13 | 1 | -5 | -1 | 2 | 2 | 5/2 | 1/2 | 1/2 | -1 | 1 | 1 | -5 | -1 | ok |
| 13 | | -1 | -5 | 1 | 2 | 2 | 5/2 | -1/2 | -1/2 | 1 | 1 | -1 | -5 | 1 | ok |
| | 14 | -1 | 5 | -1 | 2 | 2 | 5/2 | -1/2 | -1/2 | -1 | 1 | -1 | 5 | -1 | ok |
| 14 | | -5 | -1 | 1 | 2 | 2 | 5/2 | -5/2 | -1/2 | 1 | 1 | -5 | -1 | 1 | ok |
| | 15 | 3 | 1 | 3 | 2 | 1 | 3/2 | -3/2 | -1/2 | -1 | 1 | 3 | 1 | 3 | ok |
| 15 | | -1 | 3 | -3 | 2 | 1 | 3/2 | -1/2 | -1/2 | 1 | 1 | -1 | 3 | -3 | ok |
| | 16 | 1 | 3 | 3 | 2 | 1 | 3/2 | 1/2 | 1/2 | -1 | 1 | 1 | 3 | 3 | ok |
| 16 | | 1 | -3 | -3 | 2 | 1 | 3/2 | 1/2 | 1/2 | 1 | 1 | 1 | -3 | -3 | ok |
| | 17 | -1 | -3 | 3 | 2 | 1 | 3/2 | -1/2 | -1/2 | -1 | 1 | -1 | -3 | 3 | ok |
| 17 | | 3 | -1 | -3 | 2 | 1 | 3/2 | -3/2 | -1/2 | 1 | 1 | 3 | -1 | -3 | ok |
| | 18 | -3 | -1 | 3 | 2 | 1 | 3/2 | 3/2 | 1/2 | -1 | 1 | -3 | -1 | 3 | ok |
| 18 | | -3 | 1 | -3 | 2 | 1 | 3/2 | 3/2 | 1/2 | 1 | 1 | -3 | 1 | -3 | ok |
| | 19 | -1 | 1 | -5 | 2 | 0 | 1/2 | -1/2 | -1/2 | -1 | 1 | -1 | 1 | -5 | ok |
| 19 | | 1 | 1 | 5 | 2 | 0 | 1/2 | 1/2 | 1/2 | 1 | 1 | 1 | 1 | 5 | ok |
| | 20 | 1 | -1 | -5 | 2 | 0 | 1/2 | 1/2 | 1/2 | -1 | 1 | 1 | -1 | -5 | ok |
| 20 | | -1 | -1 | 5 | 2 | 0 | 1/2 | -1/2 | -1/2 | 1 | 1 | -1 | -1 | 5 | ok |
| Proton No. | Neutron No. | X | Y | Z | *n* | *L* | *j* | *m* | *s* | *i* | *parity* | X | Y | Z | Check |

Figure 3: Snapshot of a spreadsheet in which nucleon quantum values (Columns F~L) are calculated from the lattice coordinates (Columns C~E) using Eqs. 17~23. Conversely, Columns M~O are the lattice coordinates calculated from the nucleon quantum numbers using Eqs. 14~16, and Column P is a check on the equivalence between Columns C~E and M~O. Their identity ("ok") means that there is a one-to-one mapping of the lattice coordinates of each nucleon to its quantum numbers. ***That isomorphism is indeed the central claim of the lattice model: all IPM states have unique positions in the antiferromagnetic fcc lattice with isospin layering and, vice versa, every position in coordinate space (the lattice sites) corresponds to a unique quantum mechanical (IPM) state.***

Columns A and B are lists of the *approximate* empirical build-up sequence of protons and neutrons. In detail, the sequence is known to vary substantially, isotope-by-isotope. Not only does the proton or neutron sequence differ by element, but the addition of further neutrons can affect the proton sequence, and vice versa. This known variability ("configuration-mixing," "intruder states," etc.) implies that nuclear properties cannot be calculated (in either the conventional IPM or the lattice version of the IPM) using an inflexible nucleon build-up methodology. Instead, the locations of the nucleons in each isotope must be shuffled to positions that reproduce the known ***J***-value and parity ($J^{\pi}$), while minimizing Coulomb repulsion among the protons and maximizing the total number of nucleon-nucleon bonds. Such "shuffling" is presumably a consequence of local effects of the nuclear force acting among nucleons, but remains to be quantitatively demonstrated. It is noteworthy that appropriate $J^{\pi}$ lattice structures can ***always*** be constructed (if the conventional IPM can describe the quantal state of a nucleus, so can the lattice IPM), but the full set of lattice coordinates that reproduce the experimental $J^{\pi}$ is only rarely unique. As a consequence, "optimal" structures must be determined on the basis of the summation of nuclear binding effects using nuclear force coefficients, such as shown in Tables 3~6.

The encircled region shows all of the quantum characteristics of the first 40 nucleons. Just as there are no two nucleons with identical Cartesian coordinates, there are no two nucleons with the same set of quantum numbers. All further nucleon shells can be similarly defined in the lattice, as shown in a spreadsheet (using Excel functions) for calculating nuclear lattice properties. The spreadsheet can be downloaded at: www.res.kutc.kansai-u.ac.jp/~cook/QuantumNumbers.xls.